# Intra-Gauge Rotated Vector Sum (IG-RVS) for Rayleigh Fading Mitigation in Coherent φ-OTDR Systems

Igor Koltchanov[(1)], André Richter[(1)]

[(1)] VPIphotonics, Berlin, Germany, igor.koltchanov@vpiphotonics.com

**Abstract** *We propose Intra-Gauge Rotated Vector Sum (IG-RVS), a DSP-based fading mitigation method for coherent φ-OTDR. IG-RVS exploits spatial diversity within the gauge length by phase-aligning and coherently summing neighboring bins, thereby suppressing Rayleigh fading while preserving spatial resolution.: ©2026 The Author(s)*

## Introduction

The objective of phase-sensitive optical time-domain reflectometry (φ-OTDR) in distributed acoustic sensing (DAS) is to determine the spatial and temporal strain profile along the fiber [1,2]. Coherent detection of the backscattered optical field provides the *I* and *Q* quadratures from which the phase can be extracted (Fig. 1) [3]. Here, we assume baseband detection, i.e., any intrinsic acousto-optic modulator (AOM) frequency shift is compensated, or an electro-optic modulator (EOM) is used. Strain induces a phase gradient along the spatial coordinate, which is evaluated for each OTDR trace as a differential phase calculated over a distance equal to the gauge length [4,5].

After subtracting the initial phase reference, the differential phase increments between consecutive traces are integrated to reconstruct the strain evolution. To avoid phase ambiguity, phase unwrapping is applied along the slow-time axis (between traces) for each spatial bin. However, the unwrapping algorithm is highly sensitive to signal fading events caused by stochastic Rayleigh scattering [1]. This leads to large phase errors that propagate along the integration axis, degrading the reconstructed strain signal.

Several strategies have been proposed to suppress or mitigate fading events in coherent φ-OTDR [1,2,6]. Many of these approaches rely on diversity techniques that exploit the statistical independence of Rayleigh backscattering under different probing conditions, thereby reducing the probability of simultaneous fading across all channels. Examples include polarization [7,8] and frequency diversity [8-10]. A Rotated Vector Sum (RVS) method has been proposed to coherently combine independent measurements at different optical frequencies [11].

However, spatial samples located within a single gauge length can also provide independent measurements of the backscattered field, and RVS can be used to aggregate them. When the separation between neighboring bins exceeds the correlation length of the backscattered optical field, the corresponding Rayleigh contributions become statistically independent. This intrinsic spatial degree of freedom has not been considered in conventional φ-OTDR processing for fading mitigation.

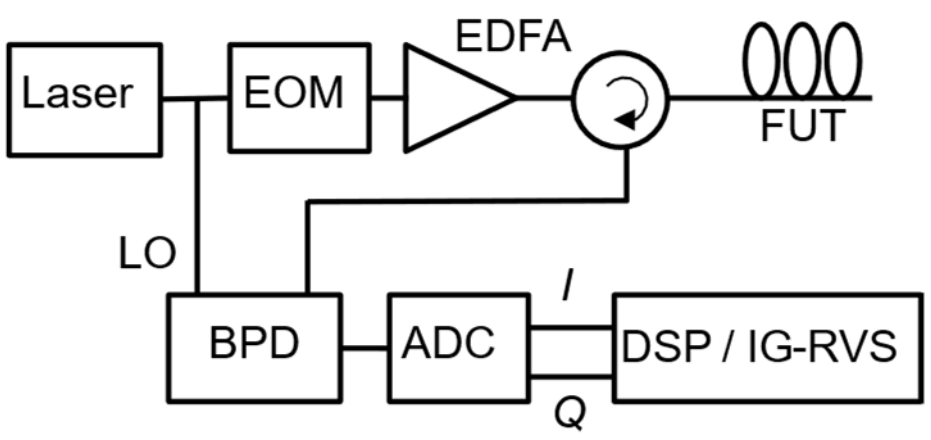

**Fig. 1:** Scheme of a φ-OTDR with IG-RVS. EOM: modulator, BPD: balanced photodetector with Local Oscillator (LO) port, ADC: Analog-Digital Converter, FUT: Fiber Under Test.

In this paper, we introduce *Intra-Gauge Rotated Vector Sum (IG-RVS)*. This DSP-based method exploits spatial diversity by phase-aligning and coherently combining neighboring bins within the gauge length. The proposed method suppresses fading and improves phase-unwrapping robustness while preserving spatial resolution without requiring additional hardware diversity. To the best of our knowledge, spatial diversity combining within the gauge length has not previously been exploited for fading mitigation in coherent φ-OTDR.

## Principle of IG-RVS

The basic idea of the RVS approach is that the phase-unstable contributions from faded channels have small magnitudes compared with those of non-faded channels and therefore contribute only weakly to the coherent sum. Hence, after an initial phase alignment, the phase evolution along the slow-time axis is mainly determined by strain variations and is therefore nearly identical for all non-faded channels. Consequently, coherent summation significantly increases the signal magnitude and effective signal-to-noise ratio.

To apply this principle to spatial channels, the selected spatial bins must be statistically independent. This requires a separation exceeding the correlation length $L_c$ of the backscattered optical field. For transform-limited probe pulses, $L_c$ is approximately equal to the pulse width. For coded OTDR schemes, $L_c$ is determined by the chip duration of the code sequence, while for chirped interrogation, it is governed by the inverse of the optical frequency sweep range.

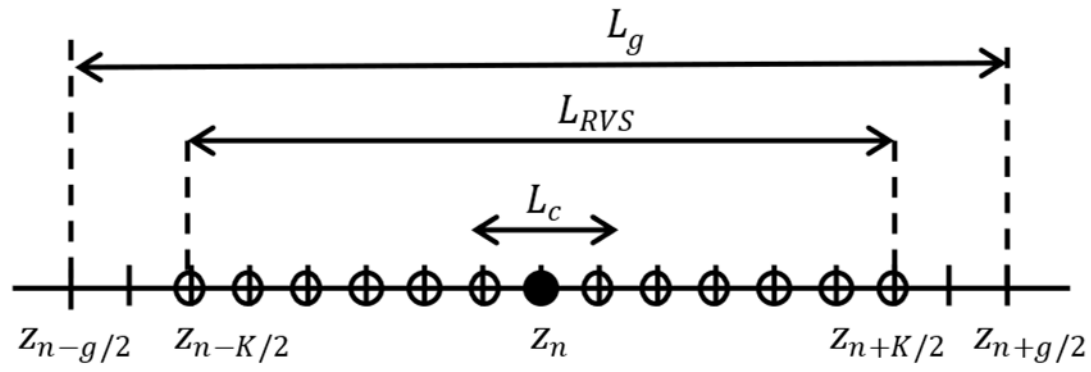

**Fig. 2**: Hierarchy of gauge length $\boldsymbol{L_g}$, RVS window $\boldsymbol{L_{RVS}}$, and coherence length $\boldsymbol{L_c}$. Dots: spatial diversity channels. Statistically independent channels are separated by more than $\boldsymbol{L_c}$.

Selecting spatial bins separated by more than the pulse width may appear to worsen spatial resolution. However, φ-OTDR evaluates the phase differential over the gauge length $L_g$, which is in practice several times larger than the pulse width. At the same time, $L_g$ is chosen short enough to satisfy the spatial resolution requirements.

Let the complex optical field reconstructed from the I/Q quadratures be

$$E(t_m, z_n) = I(t_m, z_n) + jQ(t_m, z_n)$$

where $m$ denotes the trace index (slow time) and $n$ the spatial index. The spatial coordinate is related to the ADC sampling rate $f_s$ as $z_n = c\, n/(2n_g f_s)$ where $c$ is the speed of light in vacuum and $n_g$ is the fiber group refractive index. Accordingly, the number of samples within $L_g$ is $g = 2L_g n_g f_s/c$. The spatial diversity channels are formed from the set of spatial bins located within an IG-RVS window centered at $z_n$ (Fig. 2),

$$\{z_{n-K/2}, \ldots, z_{n-1}, z_n, z_{n+1}, \ldots, z_{n+K/2}\}$$

where the window length $L_{RVS} = z_{n+K/2} - z_{n-K/2}$ is chosen such as $L_c < L_{RVS} \lesssim L_g$. Within this window, samples separated by more than $L_c$ provide statistically independent Rayleigh contributions. Consequently, the number of effectively independent spatial channels is $N \approx L_{RVS}/L_c$. The maximum number of available channels is $N_{max} \approx L_g/L_c$. For typical DAS parameters, $N_{max}$ can reach values on the order of ten, providing significantly more diversity channels than polarization-based RVS. Achieving comparable diversity using wavelength multiplexing would require substantially higher hardware complexity.

The IG-RVS processing chain consists of the following steps.

1. Differential signal calculation:
   $D(t_m, z_n) = E(t_m, z_{n+g/2})E^*(t_m, z_{n-g/2})$
2. Initial phase alignment (rotation):
   $\widetilde{D}(t_m, z_n) =$
   $D(t_m, z_n)\exp[-j\arg(D(t_0, z_n))]$
3. Spatial diversity combining (vector-sum):
   $V(t_m, z_n) = \sum_{k=-K/2}^{K/2} \widetilde{D}(t_m, z_{n+k}).$
4. Phase extraction:
   $\Psi(t_m, z_n) = \arg(V(t_m, z_n))$
5. Phase unwrapping:
   $\Phi(t_m, z_n) = \mathrm{unwrap}[\Psi(t_m, z_n)]$ along $t_m$
6. Strain reconstruction:
   $\varepsilon(t_m, z_n) = K_s/L_G \times \Phi(t_m, z_n)$

where the phase-to-strain conversion factor $K_s = \lambda/4\pi n_f(1 - p_e)$ depends on wavelength $\lambda$, fiber refractive index $n_f$, and effective strain-optic coefficient $p_e$ [1]. For $\lambda = 1550\ nm$, $n_f = 1.46$ and $p_e = 0.22$, $K_s \approx 1.1 \times 10^{-7}\mathrm{m/rad}$.

In practice, IG-RVS is implemented using a sliding spatial window. Step 2 is essential: without initial phase alignment, the differential signals $D(t_m, z_n)$ have random phases and would partially cancel during vector summation. The vector-sum operation (step 3) may optionally employ weighting coefficients and does not significantly increase the DSP complexity, since it involves only a short complex vector sum over neighboring bins. If the vector-sum operation contains only a single term (i.e., $V(t_m, z_n) = \widetilde{D}(t_m, z_n)$) IG-RVS reduces to the conventional φ-OTDR DSP chain without spatial diversity combining.

Since IG-RVS operates on the same complex differential signal as the conventional processing chain, it can be readily combined with existing DSP methods. In particular, additional spatial smoothing of the reconstructed phase along the $z$-axis may be applied, and advanced phase-unwrapping techniques such as differential unwrapping integration (DUI) [13] can be used along the slow-time axis. The method is also compatible with coded or chirped interrogation schemes, where it can be applied after the matched-filter (correlation) stage that reconstructs the complex spatial reflectogram.

**Validation by Numerical Simulation**

We validated the IG-RVS method using a sensing fiber model available in *VPItransmissionMaker™ Optical Systems* v11.7 [14]. Rayleigh backscattering is described by a frequency-domain transfer function, following the approach of [12]:

$$H(t_m, \omega) = \sum_i r_i \exp\left(-\alpha z_i - 2j\beta(\omega) z_i - \frac{j}{K_s}\int_0^{z_i} \varepsilon(t_m, s)\,ds\right)$$

where $\omega$ is the angular optical frequency, and the slow time $t_m = m/f_r$ is defined by the trace index $m$ and the pulse repetition rate $f_r$. The coordinate $z_i = i\Delta z$ denotes the numerical discretization used for Rayleigh scattering simulation and should not be confused with the spatial samples $z_n$ obtained from the ADC sampling grid. The step size $\Delta z$ is chosen to be smaller than both the coherence length $L_c$ and the probe pulse width.

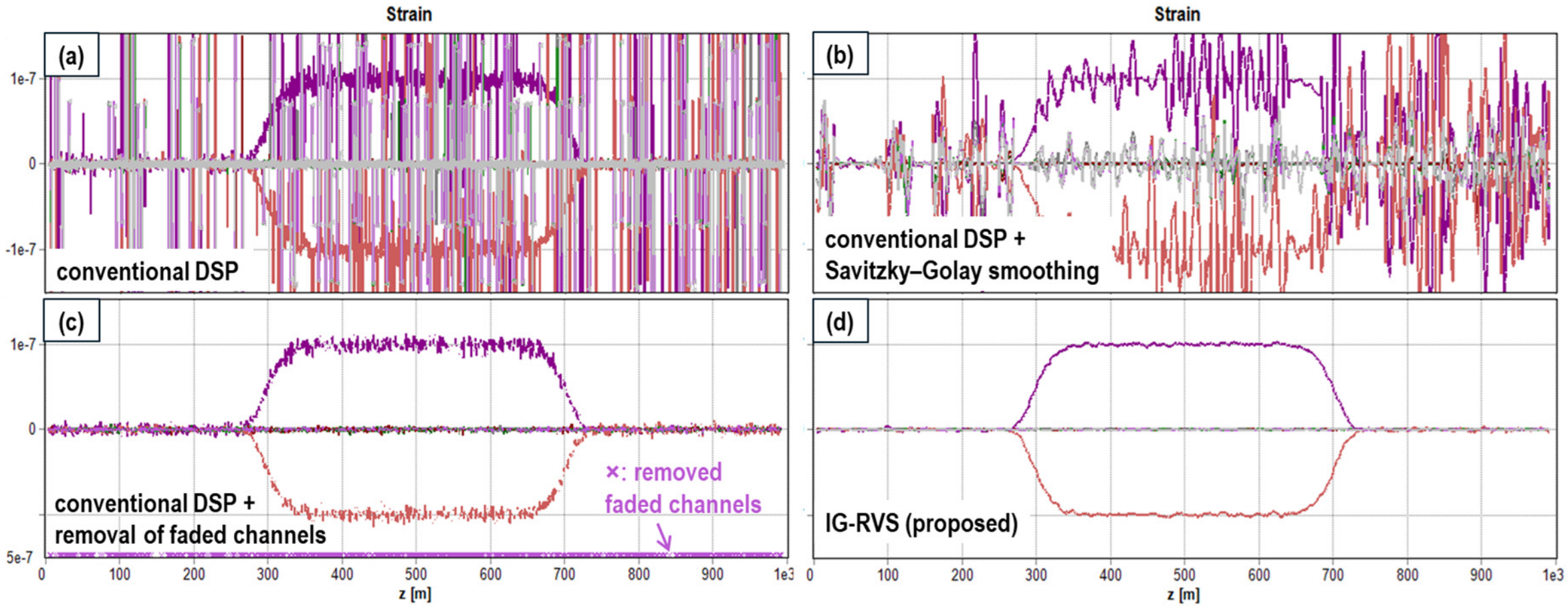

**Fig. 3**: Reconstructed strain profiles for $\boldsymbol{\varepsilon_{pk}}$ = $10^{-7}$. (a) conventional DSP; (b) conventional DSP with Savitzky–Golay spatial smoothing; (c) conventional DSP with removal of faded channels (10% quantile); (d) IG-RVS processing.

The complex coefficient $r_i$ represents the effective reflectance of Rayleigh scatterers within the numerical step $\Delta z = z_{i+1} - z_i$. It is sampled from a complex normal distribution with zero mean and variance determined by the Rayleigh backscattering coefficient $\rho$. The exponential term describes the round-trip propagation to the scatterer at position $z_i$. Here, $\alpha$ is the power attenuation coefficient; the factor of two associated with the round-trip propagation is absorbed into this term. The propagation constant $\beta(\omega)$ describes the phase delay, group delay, and chromatic dispersion, and $\varepsilon(t_m, z)$ denotes the strain distribution along the fiber.

We set in the simulations the fiber length to 1 km, the Rayleigh backscattering coefficient to -75dB/m, and the discretization step $\Delta z$ = 0.25 m. A sample strain distribution was modeled in the spatial dimension as a 6th-order super-Gaussian profile centered at 500 m with FWHM of 400 m. In the temporal dimension, it was modeled by one period of a sine wave starting at $t$ = 0.1 s, with an acoustic frequency of 10 Hz and variable strain amplitude ranging from $\varepsilon_{pk} = 10^{-8}$ to $10^{-6}$.

The transmitter generated 2000 super-Gaussian probe pulses (3rd order) at 1550 nm with peak power 200 mW, repetition rate 10 kHz, and pulse duration 10 ns. The receiver implemented a standard signal chain, including an EDFA (gain 35 dB, noise figure 3.0 dB), a photodiode (responsivity 1 A/W), a TIA with transimpedance 40 kΩ and input-referred current noise density of 10 pA/√Hz, a 5th-order Bessel low-pass filter with cut-off frequency 80 MHz, and a 14-bit ADC with sampling rate 250 MHz. In the DSP, both the gauge length and the IG-RVS window were set to 10 m, yielding approximately 10 independent spatial diversity channels for 10 ns probe pulses, consistent with the estimate discussed above.

The reconstructed strain profiles are shown in Fig. 3 for $\varepsilon_{pk} = 10^{-7}$ using four DSP options: conventional processing, conventional processing with Savitzky–Golay averaging (2nd order) applied to the reconstructed strain over a spatial window equal to the gauge length, conventional processing with removal of faded channels (10% quantile), and IG-RVS. Different trace colors correspond to time instants separated by 25 ms. In conventional processing, Rayleigh fading manifests itself as pronounced spikes. Spatial smoothing reduces high-frequency noise fluctuations but does not suppress fading-induced spikes. Removal of faded channels reduces large phase errors. However, it leads to data loss and still exhibits substantially larger fluctuations than IG-RVS. In contrast, IG-RVS suppresses fading events directly at the complex signal level, before phase extraction. As a result, the reconstructed strain profile becomes significantly smoother and more stable across the entire sensing fiber while preserving the spatial structure of the strain distribution. The same behavior is observed over the full range of strain amplitudes $\varepsilon_{pk}$ considered.

## Conclusion

We proposed applying the Intra-Gauge Rotated Vector Sum (IG-RVS) method to mitigate Rayleigh fading in coherent φ-OTDR systems. The method exploits the naturally available spatial diversity within the gauge length by coherently combining neighboring spatial bins after initial phase alignment. Numerical simulations demonstrate significant suppression of fading-induced artifacts and improved robustness of strain reconstruction. Since IG-RVS operates entirely in the DSP domain and requires no additional hardware, it can be readily integrated into existing φ-OTDR systems.

## Acknowledgement

This work was supported by the Horizon Europe Framework Programme under Grant Agreement No. 101189703 (ICON Project).